**Electron transfer and ionic displacements at the origin of the 2D electron gas at the LAO/STO interface: Direct measurements with atomic-column spatial resolution.**

By *C. Cantoni\*, J. Gazquez, F. Miletto Granozio, M. P. Oxley, M. Varela, A. R. Lupini, S. J. Pennycook, C. Aruta, U. Scotti di Uccio, P. Perna, D. Maccariello.*

[*]    Dr. C. Cantoni Corresponding-Author, Dr. J. Gazquez, Dr. M. P. Oxley, Dr. M. Varela, Dr. A. R. Lupini, and Dr. S. J. Pennycook
Materials Science and Technology Division, Oak Ridge National Laboratory, Oak Ridge TN 37831-6116 (USA)
E-mail: (cantonic@ornl.gov)
         Dr. J. Gazquez
Universidad Complutense de Madrid, Madrid 28040, Spain
         Dr. F. Miletto Granozio, Dr. C. Aruta, Prof. U. Scotti di Uccio, Dr. P. Perna, and Dr. D. Maccariello
Dipartimento di Scienze Fisiche and CNR-SPIN, Universita` di Napoli "Federico II", Compl. Univ. di Monte S. Angelo, Via Cintia, I-80126 Napoli, Italy

Interfaces between correlated oxides are currently among the most investigated systems in condensed matter. The atomic-scale engineering of such interfaces holds a great promise: the ability to manipulate electrons by locally modifying orbitals shape and occupation, yielding new quantum states that can be exploited in novel electronic devices. The interface between LAO and STO, which hosts a high mobility two-dimensional electron gas (2DEG) in spite of the large band gap of its bulk constituents, is among the most intensely debated worldwide.[1,2] After seven years of sustained theoretical and experimental work, there is no consensus regarding the dominant mechanism responsible for electrical conductivity, superconductivity, and magnetism in this system.[3] Understanding the relevant physics of the LAO/STO interface is imperative not only for the practical development of reliable devices, but also for its implications on the entire field of oxide superlattices and oxide electronics.[4] First principle calculations predict an electronic reconstruction in response to the diverging electrostatic energy generated when depositing a polar material (LAO, with alternating charged planes $(LaO)^{+1}$ $(AlO_2)^{-1}$) on a non-polar substrate (STO, with neutral planes $(SrO)^0$ $(TiO_2)^0$). In this scenario, half an electron per two-dimensional unit cell is transferred across the interface reducing the interfacial Ti valence from 4+ to 3.5+.[5-9] However, theoretical studies presuppose an atomically abrupt interface with negligible defects and/or disorder. Is such an interface realizable with current film deposition techniques? Oxygen vacancies and cation substitution are known to easily form in perovskites, acting as electron donors and giving rise to electrical conduction. Can defect-generated electrons explain all the observed properties of LAO/STO? Although simple to formulate, these questions have remained to date unanswered, partly because of sample variations due to different growth and process history, and partly because of the challenge involved with direct local probing of charge transfer, cation, and oxygen vacancies distributions with atomic-scale spatial resolution. One experimental technique suitable for these measurements is electron energy loss spectroscopy (EELS) performed in an aberration-corrected scanning transmission electron microscope (STEM), which permits atomic-scale resolved mapping of Ti, La, and O core loss spectra. However, the few STEM/EELS studies reported so far on n-type LAO/STO interfaces have been inconclusive because carried out on interfaces with high levels of oxygen deficiency and/or cation intermixing.[10-12] Here we present evidence that nearly perfect LAO/STO interfaces with negligible cation intermixing can be fabricated by pulsed laser deposition (PLD). In these samples the interfacial charge density decays over ~ 3 unit cells within the STO bulk and is not generated by oxygen vacancies concentrated at the interface. Samples with LAO thickness of 4-5 u.c. show an interfacial charge density less than 0.5 $e$ per areal unit cell ($S$) and a residual polarization of the LAO film. The direction and spatial dependence of the polarization allows us to conclude that the positive charge left behind by the electrons migrated to the interface is located towards the surface of the LAO film. These new findings rule out donor defects in STO as the main source of interfacial conductivity and offer unprecedented experimental evidence that in atomically abrupt LAO/STO interfaces the 2DEG is generated by an electronic reconstruction mechanism.

*Experimental*

We have carried out a systematic growth optimization of our PLD LAO/STO samples based on *in situ* RHEED oscillations, *in situ* plume spectroscopy, x-ray diffraction, transmission electron microscopy, field- and temperature-dependent transport properties, and optical spectroscopy, as reported in previous publications [13-16]. Details of the film fabrication process are reported in the Supplemental Information [17]. We individuated $10^{-3}$ mbar as an optimal growth oxygen pressure, corresponding to the higher end of the usually employed oxygen pressure range [18]. Optimally grown samples showed a charge density close to the theoretically expected value and a threshold for conductivity of 4 u. c. STEM cross sectional samples were prepared by precision machine polishing and low voltage Ar+ ion milling, avoiding any contact with moisture.

After transport characterization the samples were investigated by STEM/EELS focusing on a LAO thickness in the range 5 to 10 u.c. EELS shows that the Ti valence is reduced at the interface. The fraction of $Ti^{3+}$ we find cannot be accounted for by oxygen vacancies located at the interface or cation intermixing, as shown below. In addition, direct interatomic distance measurements reveal distortions of the A, B, and oxygen sublattices within the LAO film and in the STO planes close to the interface, which have a direct relationship with the density of compensating carriers. [19] Within the 5 u. c. LAO film, the net lattice polarization *P* associated with these displacements is smaller than the natural polarization of LAO, $P_o = 0.5$ *e*/*S* determined by the alternating charged planes. This shows that the electrostatic potential within the film is partially screened by a counteracting dipole Δ*P* originating from lattice distortions, therefore requiring an interfacial charge less than 0.5 *e* to compensate the polar discontinuity, as we measure.

The abruptness of the interface is a crucial aspect because even a small cation interdiffusion on the order of ~ 10% La substitution on the Sr sites can potentially give rise to a significant fraction of the predicted charge density. **Fig. 1a** is a high angle annular dark field (HAADF) image of one of our typical LAO films grown on a $TiO_2$ terminated STO single crystal by pulsed laser deposition (PLD) at an oxygen partial pressure of $10^{-3}$ mbar. **Fig. 1b** shows normalized, integrated intensity profiles for the La-$M_{4,5}$ and Ti-$L_{2,3}$ edges across the interface. The experimental data is compared in the same figure with simulated profiles from inelastic scattering modeling for an atomically abrupt LAO/STO interface. [17] In order to improve the sensitivity to any interdiffusion, the experimental profiles were averaged over a four-unit-cell-wide spectrum image. Small differences between simulation and experiment are expected due to experimental unknowns not included in the simulation such as small specimen thickness variations and drift, possible damage by ion mill, detailed aberrations, temporal incoherence etc. In this case, the discrepancy is largely attributable to a sample drift of ~ ½ u.c. over the duration of the SI (or $2\times10^{-13}$ m/s), which causes the tail of the La(Ti) signal to fall slightly above(below) the simulated profile. Overall, the experimental profiles are consistent with inelastic image simulations of an abrupt interface, and provide no evidence for significant La diffusion in the STO substrate. This finding agrees with measurements of plasma kinetics,[15] which suggest that La is not implanted into STO during growth, showing substantial differences with reports in Refs. 10 and 11. To evaluate the presence of oxygen vacancies at the interface we calculated the relative O concentration using the procedure described by Egerton.[20] As shown in **Fig. 1c**, the experimental O relative concentration shows atomic cell resolution, peaking at each $BO_2$ plane just as predicted by the simulated EELS O-K integrated intensity shown in the same figure.[21] It is important to compare the simulated O-K integrated intensity with the experimental O relative concentration as opposed to the experimental integrated intensity because the simulation does not account for the O-K fine structure. Due to differences in fine structure, the experimental O-K integrated intensity in LAO is different than in STO, independently on the actual oxygen content in the two materials. The O relative composition is constant within the analyzed region, suggesting no interfacial oxygen deficiency. This is confirmed by analysis of fine structure, discussed next.

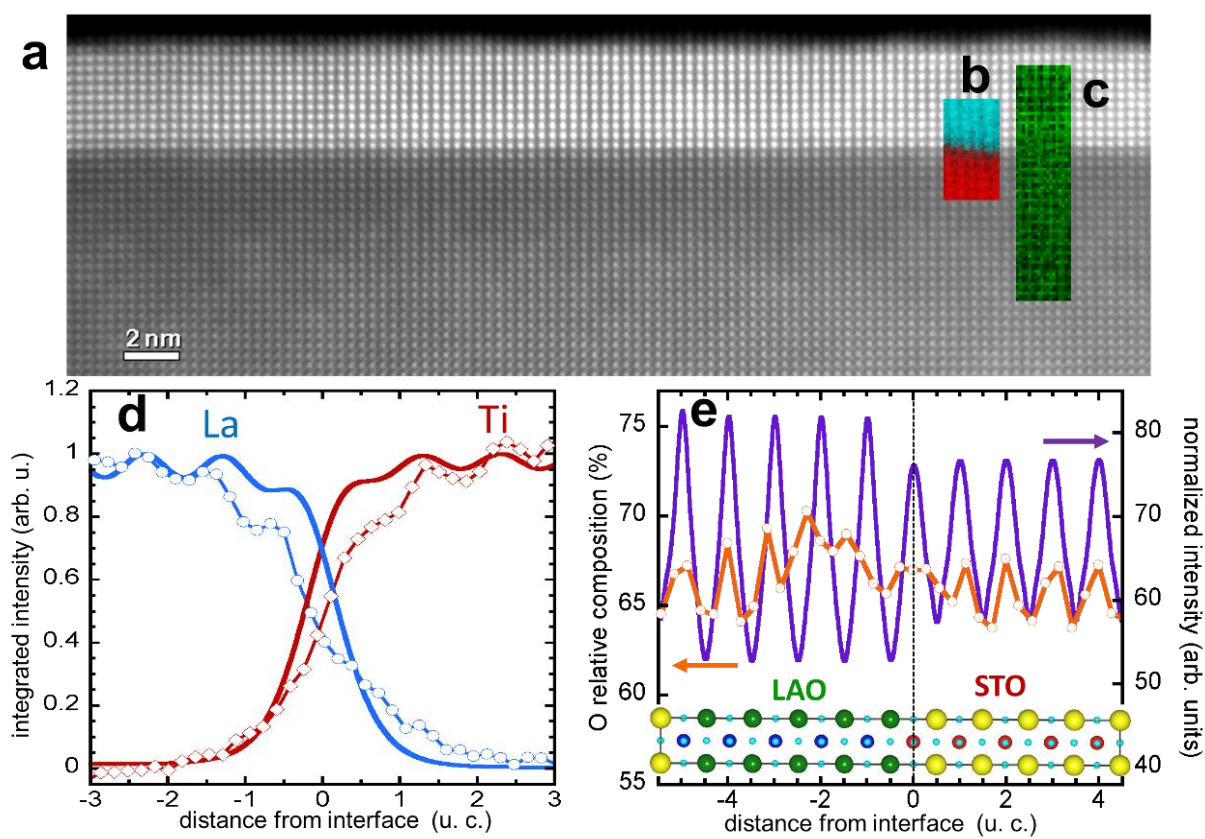

**Fig. 1**

**Figure 2a** shows a comparison of experimental Ti-$L_{2,3}$ EEL spectra for bulk STO (reference), STO well away from the interface, and the STO/LAO interface. Peaks *a* and *c* are significantly suppressed at the interface as expected for a valence change. Peak *b* is also suppressed and broadened, suggesting a possible distortion of the interfacial Ti-O octahedron.[22] The $t_{2g}/e_g$ ratio between the amplitudes of peak *c* and *d* in the Ti-$L_2$ was used to estimate the Ti valence.[23, 24] Additional insight was obtained from the O-K edge, in which the information relative to valence and lattice distortions is better separated.[25] Shown in **Fig. 2b** is a comparison between the O-K energy-loss near-edge spectrum (ELNES) well away from the interface and the O-K ELNES at the interface. The distance (ΔE) between peak A and peak C has been recognized as an accurate indicator of valence change in perovskites,[26, 27] with negligible contribution from atomic distortions.[22, 28, 29] Suppression of peak A and decrease of $\Delta E$ are evident in the graph and are unambiguously related to a decrease in the Ti valence. As an additional confirmation of the minor role of oxygen deficiency, we can compare Fig. 2b with the O-K ELNES reported in Refs. [30] and [31] for oxygen deficient $SrTiO_{3-\delta}$. For $\delta = 0.25$ the O-K ELNES is drastically different from our interface O edge, which indicates that O vacancies cannot be the only source of electrons (if each oxygen vacancy contributes 2 $e^-$, $\delta = 0.25$ corresponds to 0.5 $e^-$ transferred). Moreover, marked differences exist between the $\delta =0.13$ edge in ref. [31] and our interface edge. Quantitative analysis (see Supplemental Information[17]) shows that our oxygen content is $3.00 \pm 0.06$.

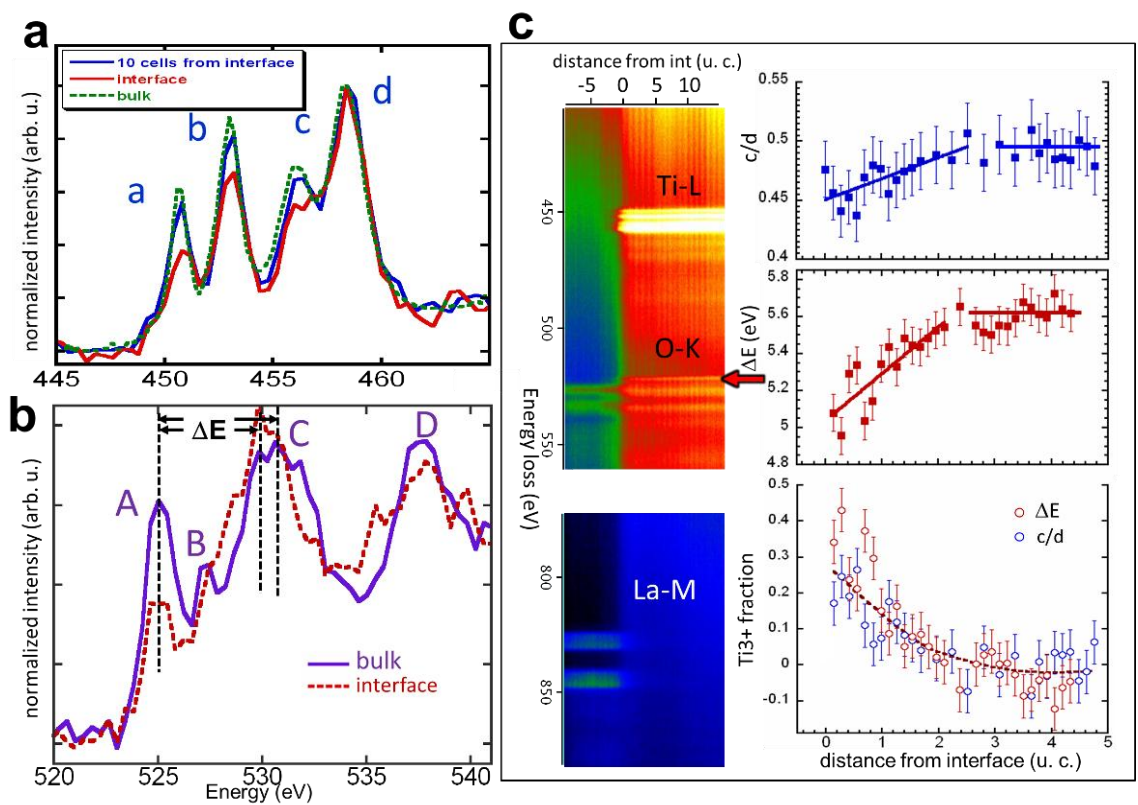

**Fig. 2**

Therefore, interfacial oxygen vacancies could only account for 0.12 $e^-$ transferred at most. **Figure 2c** shows a set of data used to extract the interfacial $Ti^{3+}$ fraction for a sample with ~ 5-u.c.-thick LAO. The first panel shows a collection of EEL spectra across the interface. The second and third panels show the measured ΔE and the ratio $c/d$ obtained from Gaussian fit of the normalized Ti-$L_2$ edge, as a function of distance from the interface. From reference spectra[32] of STO ($Ti^{4+}$), LTO ($Ti^{3+}$), and TiO ($Ti^{2+}$) we derive two linear equations for the Ti valence as function of ΔE and $c/d$ from which we extract two independent estimates of the $Ti^{3+}$ fraction as plotted in the last panel of Fig. 2c. By integrating the $Ti^{3+}$ fraction as a function of position near the interface we obtain the total interfacial charge density, which amounts to − 0.34 ± 0.09 $e/S$ and − 0.23 ± 0.09 $e/S$ for the ΔE and $c/d$ methods, respectively. The $Ti^{3+}$ fraction calculated with the same procedure for the 10 u.c.-thick sample (not shown) is 0.6 ± 0.01. While the measured injected surface charge is consistent with 0.5 $e/S$ for the 10-u.c.-thick sample, the net injected surface charge for the 5-u.c.-thick sample is smaller than the value predicted by the ionic model.[33] In presence of an incomplete charge transfer, we expect that the electric potential within the LAO is not completely screened by injected free charge. Therefore, such potential must stimulate a dielectric response within the film and be partially screened through ionic relaxations. A*b initio* computations[19, 7, 34] have shown that a dielectric LAO polarization retards charge transfer in thin films. This might correlate well with the small thickness, 5 u.c., of the thinner samples measured.

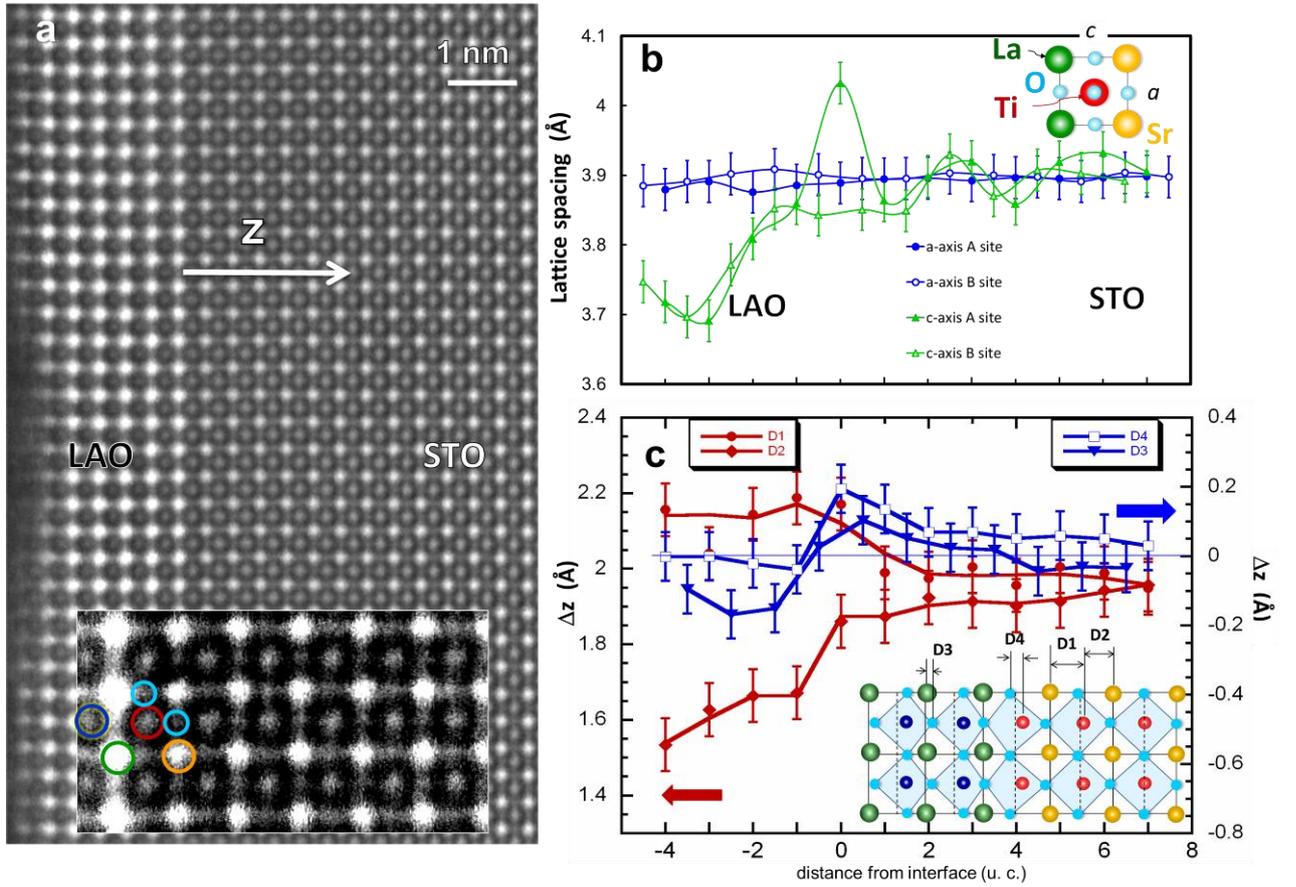

**Fig. 3**

**Figure 3a** is an HAADF image of the ~ 5-u.c.-thick sample resolving both cations and oxygen columns, acquired using a Nion UltraSTEM operating at 100 kV. Shown in **Fig. 3b and 3c** are the results of our column displacement analysis of Fig. 3a. Figure 3b shows a plot of the *a* and *c* lattice parameters obtained by averaging the distances between A-site and B-site columns. The *a* parameter calculated from either the A or B sublattices is constant across the interface, as expected for a fully strained LAO film. Remarkably, the *c* spacing calculated from A and B sublattices shows significant variations at the interface. The A sublattice appears expanded by about 4% at the interface in agreement with ref. [35]. However, in contrast to the suggestion in that work, a similar expansion is not observed in the B sublattice. Fig. 3c gives additional insight into the nature of the interfacial atomic distortions. Shown here is the distance of each B plane from its two adjacent A planes, one towards the bulk and the other towards the surface (D2 and D1 in the figure). As the interface is approached, each B plane moves closer to the A plane towards the bulk, with the distance AB remaining unchanged within the STO bulk. The displacement of oxygen creates a buckling of the AO and $BO_2$ planes indicated by the differences between cation and oxygen positions D3 and D4 in the graph. Both D3 and D4 are close to zero in the STO bulk and increase as the interface is approached. They reach a maximum value at the interface and then change sign in the LAO film. This behavior indicates that, starting from about 3-4 unit cells from the interface, the oxygen atoms in STO move in the direction opposite to the Ti ions, that is, towards the surface. In the LAO film near to the interface the oxygen atoms move towards the bulk, in the same directions as the Al ions. The ferroelectric-like cell distortions observed in LAO and STO are consistent with results from density functional theory,[19, 36] and in part with previous structural studies.[37-41] In absence of any lattice distortions and neglecting the electronic polarization, the dipole moment of an individual LaO-AlO$_2$ unit is $d = ea/2$ ($a$ = lattice parameter; $e$ = electron charge), corresponding to an intrinsic LAO polarization $P_o = e/2S$, where $S$ is the cell surface. The corresponding polarization in STO is null. The displacements we measure are responsible for an induced polarization $\Delta P$ that contrasts the intrinsic polarization $P_o$ of the LAO film and extends for a few unit cells into the STO, partially compensating the polar discontinuity. In the LAO film the main contribution to $\Delta P$ arises from the unequal distance between La and Al planes, while in STO the main contribution is given by the buckling of the TiO$_2$ and SrO planes. The resulting polarization $\boldsymbol{P} = \boldsymbol{P_o} + \Delta \boldsymbol{P}$ calculated using formal ionic charges and the atomic position illustrated in Fig. 3c is shown in **Fig. 4**. The polarization is positive in our frame of reference, thus $\boldsymbol{P}$ points from LAO to STO and varies from an average of 0.36 ± 0.07 $e/S$ within the LAO film to zero over 3-4 STO unit cells. As expected, within LAO, $P$ is smaller than the "built in" polarization due to the alternating charged planes, $P_o$ = 0.5 $e/S$. Therefore, we are in an intermediate situation between a) the case of a thin overlayer[37, 40], for which charge injection does not take place and the electric potential is screened only by ionic displacements, and b) the case of thick overlayers for which $P$ = 0.5 $e/S$ in LAO and the electrostatic potential is completely screened by the injection of a net interface free charge of − 0.5 $e/S$. The spatial dependence of the lattice polarization offers by itself a remarkable indication of the intrinsic nature of the 2DEG. The decay of $P$ within the STO occurs on the same scale as the decay of the interfacial charge measured by EELS, and both agree well with a first-principle model of the LAO/STO interface. In fact, as demonstrated in Ref. [19], a partial compensation scenario with a polarization ≤ 0.3 $e/S$ in LAO is realized through injection of a net surface free charge of − 0.3 $e/S$ distributed over 3-4 STO unit cells, just as we measure. Moreover, the polarization is constant throughout the LAO film allowing us to draw an analogy with a parallel plate capacitor in which the negative and positive charges are distributed on planes located at the STO interface and at the surface of the LAO film respectively. The direction of the electric field, pointing from LAO to STO and vanishing in the STO bulk, as probed directly through the polarization plot in Fig. 4, shows that the positive charges left behind by the electrons that migrated to the interface cannot be located on the bulk-STO side but must be at the top of the LAO film. The nature and the exact position of the "donors" of the electronic reconstruction process is a crucial but relatively unaddressed issue, so far. Due to the heavy energetic cost of displacing charges from the largely ionic $La^{3+}Al^{3+}O^{2-}_3$ lattice, we cannot exclude that relatively free electrons, loosely bonded around positively charged O vacancies located close to the LAO surface are transferred to the STO.

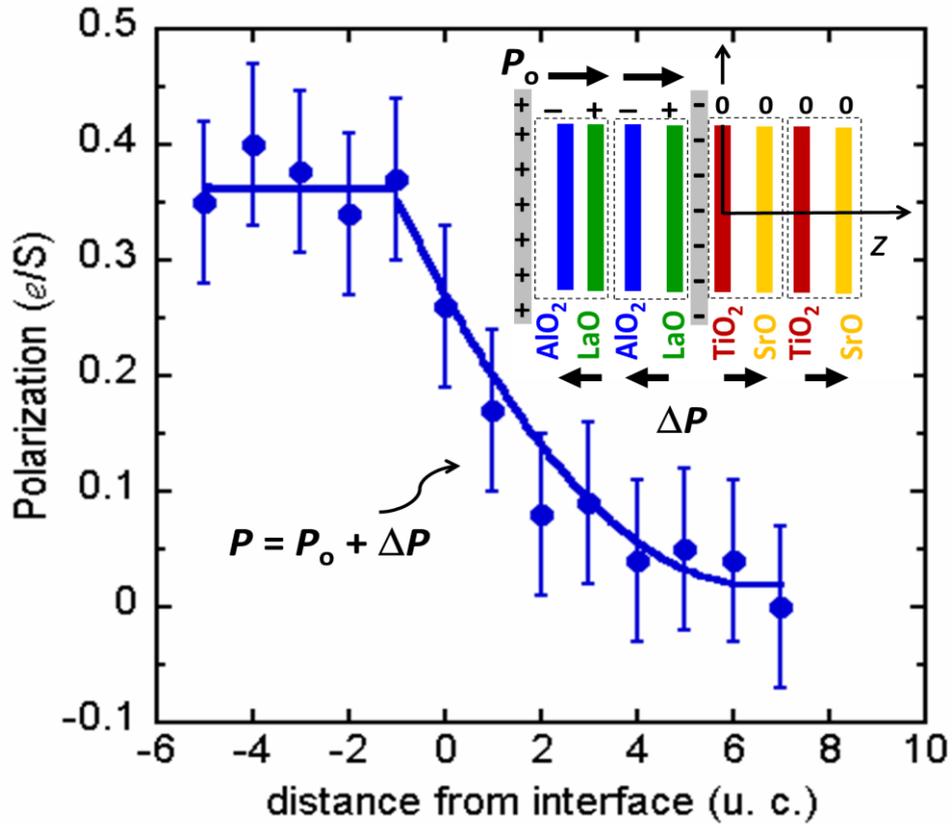

**Fig. 4**

In summary, we report direct observation of Ti valence reduction near the interface from 4 to 3.72 ± 0.09 for interfaces with a 5-u.c.-thick LAO film, and from 4 to 3.4 ± 0.1 for interfaces with a 10-u.c. LAO film. We provide evidence that charge injection is not the only reconstruction mechanism occurring to reduce the potential build-up in thin LAO overlayers. The built-in dipole in the LAO layer is also partially compensated by ionic displacements occurring in both LAO and STO. In STO, the ions are displaced over a length of about 3-4 unit cells from the interface. This picture is consistent with a partial compensation scenario as predicted by first-principle calculations, which is realized in our case for an overlayer thickness close to the critical 4-u.c.-value. Our measurements rule out a significant contribution from cation intermixing and oxygen vacancies either concentrated at the interface or distributed within the STO. The polarization profile shows that the positive charges are located towards the LAO surface, in agreement with an electronic reconstruction scenario. Overall, these experimental results identify key intrinsic phenomena for the formation of the 2DEG at the LAO/STO interface.

*Acknowledgements*

The authors are grateful to M. Stengel for helpful discussions, to J. Luck for sample preparation, and to D. Marré and A. Gadaleta for transport measurements. Research in U.S. sponsored by US Department of Energy, Office of Science, Division of Materials Sciences and Engineering and Office of Electricity Delivery and Energy Reliability. Research in Europe sponsored from European Union Seventh Framework Programme (FP7/2007-2013) under grant agreement N. 264098 - MAMA, and from the Italian Ministry of Education, University and Research (MIUR) under Grant Agreement PRIN 2008 - 2DEG FOXI. J. G. acknowledges financial support from European Research Council Starting Investigator Award.

**Figure Captions**

**Fig. 1**. Structure and chemical composition of the LAO/STO interface at the atomic scale. (a)-(c) HAADF micrograph and superimposed EELS Ti and La elemental map (b) and O map (c). (d) Experimental La-$M_{4,5}$ and Ti-$L_{2,3}$ EELS profiles from (b) (open symbols), and corresponding simulated profiles (solid lines) for an atomically abrupt interface. (e) Simulated (solid line) and experimental (line + symbols) O-K profiles from (c). The experimental curve is $N_o /(N_o + N_{Ti} + N_{La})$, with $N$ areal density. Simulation and experimental data express two slightly different quantities and their different average does not indicate different oxygen content (see text). Data in (a), (b), and (d) are from a VG 501 operated at 100 kV and equipped with a Nion aberration corrector and an Enfina EEL spectrometer. Data and in (c), and (e) are from a Nion UltraSTEM operated at 100 kV.

**Fig. 2**. Analysis of Ti-$L_{2,3}$ and O-K edges. (a) Comparison between the bulk STO Ti-$L_{2,3}$ ELNES (fine dashed line), the STO Ti-$L_{2,3}$ ELNES at 10 unit cells away from the interface (thick dashed line), and the STO Ti-$L_{2,3}$ ELNES at the interfacial $TiO_2$ plane. The last two spectra are part of the same line scan. (b) Comparison between O-K in STO well away from the interface (blue, solid line) and O-K at the interfacial $TiO_2$ plane (red, dotted line). Spectra are displayed after noise reduction by Principle Component Analysis (PCA). (c) Left panel: EELS data acquired in a line scan showing the evolution across the interface of the Ti-L, O-K, and La-M edges. Red arrow indicates peak A in O-K . Right panel: $\Delta E$ from the O-K edge as a function of distance from the interface (top). Middle graph: Evolution of $c/d$ in the T-$L_2$ edge as a function of distance from the interface. Bottom graph: $Ti^{3+}$ fraction calculated from $c/d$ (blue symbols) and $\Delta E$ data (red symbols). Data refer to the ~5-u.c.-thick sample. The curves shown in the graphs are guides to the eye.

**Fig. 3**. Atom displacements in proximity of the LAO/STO interface. (a) LAO/STO interface imaged by HAADF in a Nion UltraSTEM operated at 100 kV. The inset is an enlarged portion of the image (raw data) with color coded circles indicating the position of each atomic column (red = Ti, blue = Al, yellow = Sr, green = La, cyan = O). (b) *a*-axis spacing as a function of position for A (closed circles) and B (open circles) sublattices. *c*-axis spacing as a function of position for the A (closed triangles) and B (open triangles) sublattices. The data are averages in the direction parallel to the interface. (c) Red symbols: Distance between the B plane at a given *z* coordinate and the A plane on its left (D1), and on its right (D2). Blue symbols: Difference between cation and oxygen position in AO planes (D3) and $BO_2$ planes (D4). (c) Schematic (not in scale) of atomic displacements.

**Fig. 4.** Polarization per unit cell calculated using atomic positions from Fig. 3. The inset shows a parallel plate capacitor model discussed in the text.

## Supplemental Information

Sample fabrication

The samples analyzed in this work were deposited on nominally $TiO_2$ terminated STO substrates, chemically treated in de-ionized (DI) water and buffered-HF (BHF) according to [1] The growth was performed by Reflection High Energy Electron Diffraction (RHEED) assisted PLD (KrF excimer laser, 248nm) with a typical fluence of about $2J/cm^2$ at the target, a substrate temperature of 780°C and an oxygen pressures of $10^{-3}$ mbar. Substrate positioning at the edge of the plume was performed with the help of an intensified charge coupled device camera (ICCD), also allowing us to analyze the kinetic energy and the chemical state of the species in the plume in different deposition conditions [2,3,4], for both LAO/STO and $LaGaO_3$/STO [5] interfaces. RHEED data of a typical LAO/STO sample are reported in fig. 1. The nonlinear optical response was systematically analyzed by second harmonic generation (SHG), suggesting an increase in the polarization of the STO lattice near the interface at thickness exceeding 3 u.c. [6,7], in agreement with the atomic displacements reported in this work. The resistivity for the 10 u.c. sample discussed in the main text is shown in fig. 1b. Similar behavior is routinely observed for samples fabricated in similar conditions.

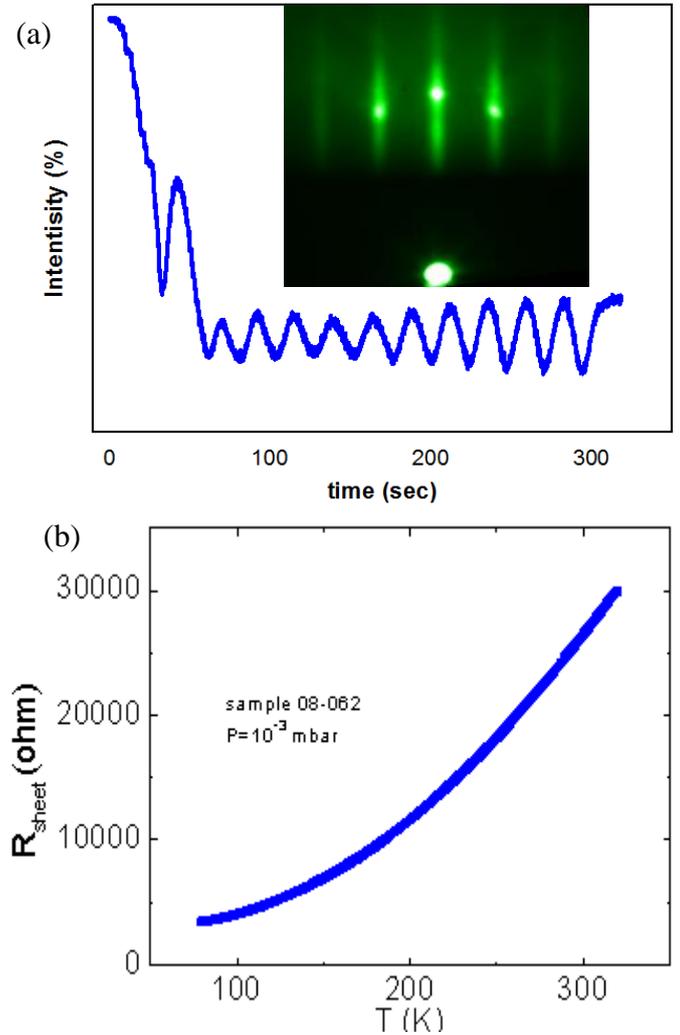

**Fig. 1.** (a) RHEED oscillations and RHEED pattern at the end of the deposition (inset) of a typical LAO/STO interface. (b) R(T) curve for a sample discussed in the main text.

Hall Effect measurements indicate a sheet carrier density in the range $1\times10^{14}$–$3\times10^{14}$ $cm^{-2}$ at room temperature with no clear dependence on the growth oxygen pressure in the range $10^{-5}$–$10^{-2}$ mbar.

EELS simulation

The EELS simulation for the Ti-$L_{2,3}$ and La $M_{4,5}$ profiles was performed using a 1×10 supercell, parameters dictated by the microscope and experimental acquisition conditions, and the TEM foil thickness (0.3 $\lambda$ in this case, where $\lambda$ is the inelastic mean free path). All calculations are based on the Bloch wave method outlined in [8]. The EELS inelastic scattering coefficients were calculated using isolated atomic models, with the bound state calculated using a relativistically corrected Hartree Foch formulation and Hartree Slater continuum states. [9].

Estimate of oxygen content

The evolution of the O-K ELNES with increasing O vacancies in STO has been well documented theoretically and experimentally in refs. [10] and [11]. Both studies show the same trends in the O-K ELNES changes as the number of O vacancies is increased. In particular, peak C, which arises from oxygen scattering, is gradually removed with increasing oxygen deficiency. In addition, peak B increases in intensity as oxygen is removed. The enhancement of peak B "provides strong evidence for the presence of O vacancies since an increase in the intensity of a peak cannot be induced by distortions since a loss of symmetry always removes the degeneracy of paths thus damping the ELNES" [10]. Multiple scattering (MS) calculations show a strong sensitivity of the amplitude of peak B to oxygen deficiency: An increase in peak intensity is already detectable with one O removed from an eight-shell STO cluster containing 99 atoms ($\delta$ = 0.05 or 1.7% O deficiency). Fig. 2 is a plot of the amplitude of peaks A and B in the experimental O-K ELNES as a function of distance from the interface obtained through peak analysis and deconvolution of the O-K edge in the spectrum image used for determining the Ti valence. The plot shows that the intensity of peak B is constant in the bulk and decreases

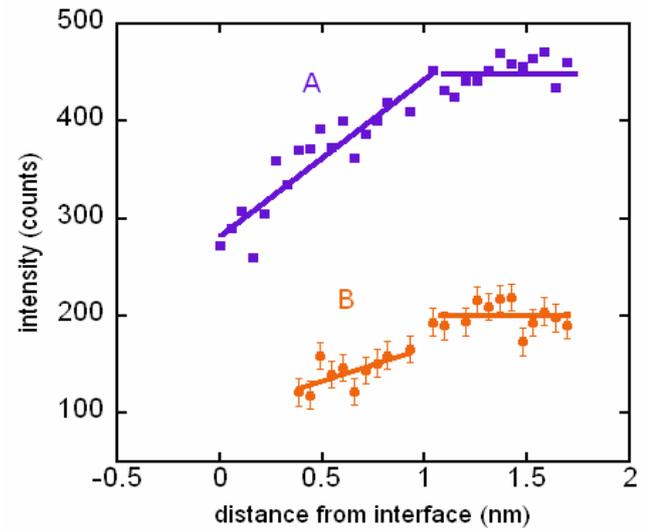

**Fig. 2.** Evolution of the amplitude of peak A and B as function of position

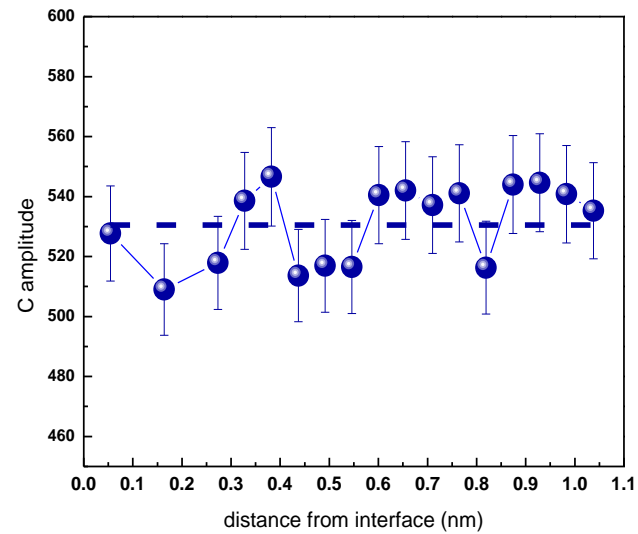

**Fig. 3.** Evolution of the amplitude of peak C as function of position

in proximity to the interface. The decrease close to the interface is caused by oversampling of the LAO film (for which peaks A and B are absent) due to beam dechanneling. According to MS simulations in ref. [8], the amplitude of peak B increases at the rate of about 4% for a 1%

O deficiency. On the basis of this observation we conclude that, if present, an oxygen deficiency as small as 2% at distances greater than 1 *nm* from the interface (Fig. 1) would change the slope and the intercept of the linear fit to the intensity of peak B to a positive value greater than the fit error. Our sensitivity to oxygen deficiency for distances ≥ 1 *nm* from the interface is therefore 2%. The absence of oxygen vacancies in the region between 0 and 1 *nm* from the interface is then inferred by the fact that there is no change in the amplitude of peak C in the two regions (see Fig. 3). The oxygen sensitivity in this case is derived from the dependence of the amplitude of peak C on oxygen content and is also 2%. These estimates provide an additional evidence for the minor role of oxygen vacancies in originating the interfacial conductivity.

Fig. 2 also shows that proximity to LAO is not sufficient to explain the observed suppression of peak A at the interface, which has to be attributed to a change in valence of the interfacial Ti atoms. In fact, peak A decreases about three times faster than peak B, which can only be explained by the presence of electrons in the $t_{2g}$ levels (but not in the $e_g$ levels).

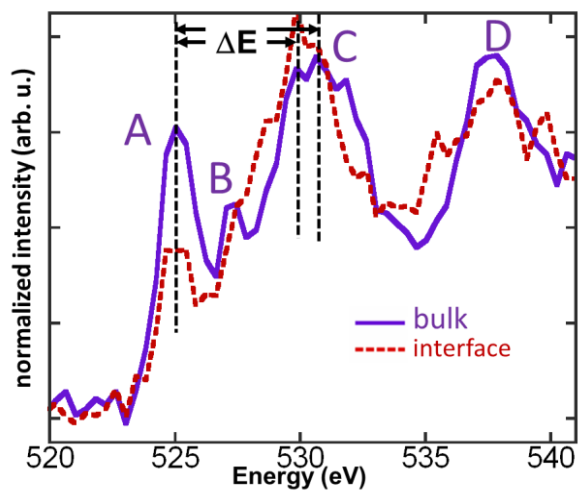

**Fig. 4**. comparison between O-K ELNES at the interface and within the STO bulk

A comparison between O-K ELNES at the interface and within the STO bulk was shown in the main text and is copied here (Fig. 4) for clarity. We could not detect any decrease in intensity of peak C or any increase in intensity of peak B. The only relevant changes are a decrease in intensity of peak A and a reduction of the distance between A and C, both attributable to Ti valence change.